\documentclass[prb,english,aps,tighten,twocolumn, superscriptaddress]{revtex4}
\usepackage[T1]{fontenc}
\usepackage[latin1]{inputenc}
\usepackage{graphicx}

\makeatletter \tolerance = 10000 \tolerance = 10000

\usepackage{color}
\usepackage{babel}

\begin{document}
\title{Tuning the electronic structures of armchair graphene nanoribbons through
chemical edge modification: A theoretical study}

\author{Z. F. Wang}
\affiliation{Hefei National Laboratory for Physical Sciences at
Microscale, University of Science and Technology of China, Hefei,
Anhui 230026, People's Republic of China}

\author{Qunxiang Li}\thanks{Corresponding author. E-mail: liqun@ustc.edu.cn}
\affiliation{Hefei National Laboratory for Physical Sciences at
Microscale, University of Science and Technology of China, Hefei,
Anhui 230026, People's Republic of China}

\author{Huaixiu Zheng}
\affiliation{Hefei National Laboratory for Physical Sciences at
Microscale, University of Science and Technology of China, Hefei,
Anhui 230026, People's Republic of China}

\author{Hao Ren}
\affiliation{Hefei National Laboratory for Physical Sciences at
Microscale, University of Science and Technology of China, Hefei,
Anhui 230026, People's Republic of China}

\author{Haibin Su}
\affiliation{School of Materials Science and Engineering, Nanyang
Technological University, 50 Nanyang Avenue, 639798, Singapore}

\author{Q. W. Shi}
\affiliation{Hefei National Laboratory for Physical Sciences at
Microscale, University of Science and Technology of China, Hefei,
Anhui 230026, People's Republic of China}

\author{Jie Chen}\thanks{Corresponding author. E-mail: jchen@ece.ualberta.ca}
\affiliation{Electrical and Computer Engineering, University of
Alberta, AB T6G 2V4, Canada}

\begin{abstract}
We report combined first-principle and tight-binding (TB) calculations to
simulate the effects of chemical edge modifications on structural
and electronic properties. The C-C bond lengths and bond angles near
the GNR edge have considerable changes when edge carbon atoms are bounded
to different atoms. By introducing a phenomenological hopping
parameter $t_{1}$ for nearest-neighboring hopping to
represent various chemical edge modifications, we investigated the
electronic structural changes of nanoribbons with different widths based on the
tight-binding scheme. Theoretical results show that addends can
change the band structures of armchair GNRs and even result in
observable metal-to-insulator transition.


\end{abstract}

\maketitle

Carbon nanotubes (CNTs) with chiral vectors $(n,m)$ are metallic
when $(2n+m)$ or equivalently $(n-m)$ is a multiple of $3$. Armchair
CNTs with chiral vectors $(n,n)$ are always metallic while the
zigzag CNTs $(n,0)$ are metallic only when $n$ is a multiple of $3$.
In general, the electronic properties can be modified by attaching
atoms or molecules along CNTs' sidewalls using chemical
methods.\cite{CPL00,SSP05,CR05} The recent development of graphene,
a sheet of an unrolled CNT, has attracted a great deal of research
attention.{\cite{Graphene-Science04, Graphene-Exp}} Graphene is a
two-dimensional one-atom-thick carbon layer with carbon atoms
arranged in a honeycomb lattice. Experiments by using the mechanical
method and the epitaxial growth method show it is possible to make
GNRs with various widths.{\cite{Graphene-Exp, CBSCI06,CBJPCB04}}
Several experiments have also been performed recently to investigate
the transport properties of graphene.{\cite{Graphene-Exp}} The
results show that graphene is an interesting conductor in which
electrons move like massless and relativistic particles. Several
groups show STM/STS graphene images with well-defined
hydrogen-terminated edges in an ultrahigh-vacuum (UHV) environment.
{\cite{YNASS05,YKPRB05, YNPRB06}}

An interesting question to investigate is whether a metallic CNT is
still metallic when it is unwrapped into a graphene nanoribbon
(GNR).{\cite{PAC90, PRB96, JPSJ96}} For simplicity, we chose the
same definition for GNRs as for CNTs in this paper.
An unwrapped $(n,m)$ CNT, for instance, is called a $(n,m)$ GNR.
Previous theoretical studies based on the tight-binding (TB)
approximation have examined the electronic states and energy
dispersion relations of GNRs with the assumption that hydrogen atoms
are attached to the carbon atoms on the GNR edge. These calculations show
that the metallic or insulating feature in GNRs are different from
that in CNTs.{\cite{YLPRB02,9,10}} Zigzag GNRs $(n,n)$, which
correspond to unwrapped armchair CNTs with zigzag edges, are still
metallic (if the spin degree of freedom is not considered). The
electronic structure of armchair $(n,0)$ GNRs depends strongly on the
nanoribbon width.{\cite{YLPRB02,9,10}} Armchair GNRs are metallic
when $n=3p+1$ ($p$ is an integer), and otherwise they are
insulating. That is to say, GNRs can be made either as metallic or
as semiconducting materials by controlling their width or chirality.
This remarkable characteristic is very useful in making
graphene-based nanoscale devices. In addition, GNRs are much easier
to manipulate than CNTs due to their flat structure, and thus they
can be tailored by using lithography and etching techniques.

Because each edge carbon atom of GNRs is only bounded to two neighboring
carbon atoms, a dangling carbon bond offers a remarkable opportunity
to manipulate the electronic properties of GNRs. This can be done by attaching
different atoms or functional molecular groups to the dangling carbon
atom. Similar to the functionalization of CNT devices
along edge dangling bonds,{\cite{SSP05,CR05, MolDevices}} the
electronic properties of GNRs can also be alternated by chemical
edge modifications. In this paper, we examine the geometric
deformation of finite-width armchair GNRs caused by different edge
addends. Our first-principle calculations show that the C-C bond
length and bond angle near the edge undergo observable changes. To
include the effect of the deformation on the graphene ribbon edge,
we introduce a phenomenological hopping parameter $t_{1}$ in our TB approximation
calculations. Our simulations show that the energy gap depends on
ribbon width and hopping parameter $t_{1}$. A nonzero energy gap
exists for $(3p+1,0)$ armchair GNRs, which means the
metal-to-insulator transition can be achieved by edge modifications.

\begin{figure}[htbp]
\includegraphics[width=7cm]{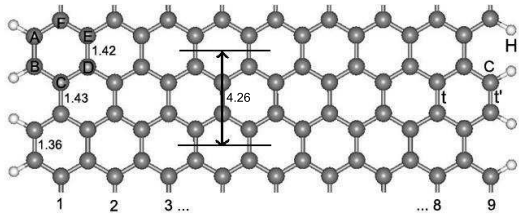}
\caption{Optimized structure of a $(9,0)$ armchair GNR (all edge
bonds are bound by hydrogen atoms). Here, the bond length is in
{\AA}. Note that a zigzag $(n,0)$ GNR  can be rolled to form an
armchair ($(n,0)$ CNT, and an armchair  $(n,n)$ GNR can be rolled as
a zigzag $(n,n)$) CNT.} \label{fig1}
\end{figure}

The increments of C-C bonds and bonding angles at the nanoribbon
edge have been reported based on the TB approximation
calculations.{\cite{PAC90, PRB96, JPSJ96}}. In this paper, we first
estimate nanoribbon geometric deformation caused by various chemical
addends. For a simple case (each armchair GNR edge carbon atom is
saturated by one hydrogen atom), we evaluate geometric and
electronic structure changes using the first-principle method. Our
optimizations employ the Vienna \textit{ab initio} simulation
package \cite{vasp1,vasp2}, which is implemented based on the local
density approximation \cite{LDA} of the density functional theory
(DFT).\cite{DFT} The electron-ion interaction is described by the
ultra-soft pseudopotentials \cite{uspp} and the energy cutoff is set
to be $286.6$ eV. The atoms' positions are optimized in order to
reach the minimum energy with the Hellmann-Feynman forces less than
$0.02$ eV/\AA. The results show that the geometric relaxation
localizes near the edge. Only the bond lengths and angles of edge
carbon atoms in armchair GNRs have considerable changes comparing
with those of ideal graphene ribbons. For example, $\angle$
BAF$=$$121.6^{\circ}$ and $\angle$BCD$=$$118.3^{\circ}$. The
interatomic distance between A and B sites (d$_{AB}$) of $(9,0)$ GNR
is reduced from 1.42 \AA\ to 1.36 \AA, and d$_{BC}$=1.40 \AA\,
d$_{CD}$=1.40 \AA\, and d$_{DE}$=1.42 \AA\, respectively, as shown
in Fig.~\ref{fig1}.

The geometric deformation of armchair GNRs, however, depends on
various kinds of chemical addends. For example, the edge carbon
atoms connect to F atoms, the edge C-C bond lengths shorten to be
1.35 {\AA}(decreasing about 5{\%}). In general, this kind of
geometric deformation results in the changes of hopping parameter
between two neighboring carbon atoms on the GNR edge. The parameter
change is defined as

\begin{equation}
\Delta t=\frac {< 2p^{\prime}_z|\hat{H_1}|2p^{\prime}_z>-<
2p_z|\hat{H_0}|2p_z>}{<2p_z|\hat{H_0}|2p_z>}.
\end{equation}

\noindent Here, $\hat{H_1}$ and $\hat{H_0}$ are the Hamiltonians of
the system with and without chemical edge modification,
respectively. $2p^{\prime}_z$ and $2p_z$ are the atomic orbitals of
the coupled neighboring carbon atoms at the edge with the optimized
bond length and with the bond length of 1.42 {\AA}, respectively.
Our calculations are conducted by using the SIESTA code in real
space and single-zeta (SZ) basis.\cite{Estimation} One can expect
that the edge hopping parameter can increase (or decrease) as edge
C-C distance is shortened (or expanded). The edge hopping parameter
change for the hydrogen-saturated case is predicted to be $10.1$
{\%} based on the DFT, which is consistent with the numerical result
based on the TB approximation.\cite{170}

In previous TB approximation calculations,{\cite{PRB96,17}} the
dangling bonds are assumed to be saturated by hydrogen atoms and
thus all transfer integrals between the nearest-neighbor sites are
set to have the same values. This simple choice of hopping
parameter, however, does not consider geometric distortions at the
nanoribbon edge. It is important to extend this existing scheme in
order to understand the impacts of chemical edge modification on
electronic properties of armchair GNRs. For simplicity, we adopt the
TB approach to study these impacts. We choose the value of $t_{1}$
either smaller or larger than the hopping parameter $t = 2.66$ eV of
inner C-C bonds to simulate different chemical addends. Our
theoretical results show that only parameter $t_{1}$ is sufficient
to describe the electronic structure changes of armchair GNRs.

\begin{figure}[htbp]
\includegraphics[width=7cm]{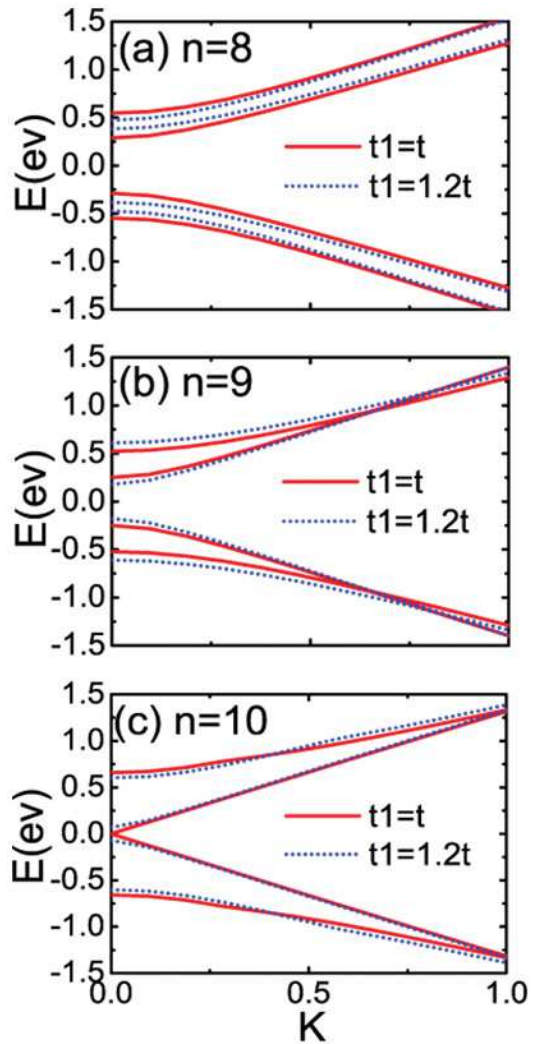}
\caption{(color online) Top two valence bands and bottom two
conduction bands of zigzag GNRs in $t_{1}=t$ and $t_{1}=1.2t$ cases.
Here $t = -2.66$ eV is the hopping parameter of two-dimensional
infinite graphene with the lattice constant $a_{c-c}=1.42\AA$.}
\end{figure}

Next, we calculate the band structures of armchair GNRs, when edge GNR
dangling bonds are bounded by atoms or molecular groups. We select
ideal armchair GNRs with indices $(n,0)$ and calculate the band
structures of three different ribbon widths ($n=8, 9$, and $10$) by
setting $t_{1}=t$.  The results are shown in Fig. 2(a) to (c) for
$n=8, 9$, and $10$ with red solid lines, respectively. For clarity,
we set the Fermi level to zero ($E_{F}=0$) and the wave number is
normalized by the primitive translation vector for each GNR. In this
study, we chose $(8,0)$, $(9,0)$ and $(10,0)$ GNRs (their
widths are around 2 nm) as examples to represent $(3p+2,0),(3p,0)$
and $(3p+1,0)$ GNRs (where $p$ is an integer; and they correspond to three
different width armchair GNRs labeled with $(3p+1,0)$, $(3p,0)$, and
$(3p+2,0)$ in the literature \cite{Louie}). It
is clear that the band structure of an armchair GNR depends on its
width, and the energy gap is $0.58$, $0.50$, and $0.0$ eV for $(8,0)$,
$(9,0)$, and $(10,0)$ armchair GNRs, respectively. Armchair GNRs $(n,0)$ are
metallic when $n=3p+1$ as shown in Fig. 2(c). The energy gaps are
nonzero for $n=3p$ and $n=3p+2$ GNR, in which GNRs become
insulating. This result matches the previous reported
results.{\cite{PRB96}} To simulate chemical edge modification
effects, $t_{1}$ is set to be $1.2t$. The $\pi$ and $\pi^{\star}$
conduction bands are no longer degenerate at $k=0$ for a  $(10,0)$
GNR and energy gap becomes $0.14$ eV, which leads to
metal-to-insulator transition. The DFT results suggest that both
quantum confinement and edge effect cause the opening of energy
gap.{\cite{Louie}} Although $(8,0)$ and $(9,0)$ GNRs remain
insulating when $t_{1}=1.2t$, the energy gaps is changed to $0.76$
eV (increased) and $0.36$ eV (decreased) by changing of the edge
hopping parameter. Obviously, these theoretical results show that
the electronic property of armchair GNRs is tunable via chemical
edge modification.

The band structure of zigzag $(n,n)$ GNRs  are similar to those of
CNTs except for the existence of edge states that are caused by the
gauge field at GNR edges.{\cite{PRB96, JPSJ96}} Due to the localized
states at the edges, the uppermost valence band and the lowest
conduction band are always degenerated at the Fermi level when $k_0
\leq|k|\leq\pi$. $k_0$ is slightly less than $\frac {2\pi}{3}$ for
finite-width zigzag GNRs. If considering the spin degree
of freedom in the DFT calculations, a hydrogen-saturated zigzag GNR is
predicted to have a magnetic insulating ground state.\cite{Louie} By
selecting different $t_{1}$, we observe that the electronic
structure near the Fermi level  changes slightly, which indicates
the electronic properties or carrier transport of zigzag GNRs are
stable and insensitive to the edge modifications.

\begin{figure}[htbp]
\includegraphics[width=7cm]{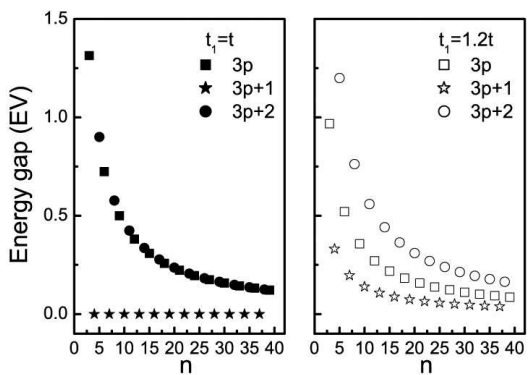}
\caption{The energy gap of armchair GNRs depends on their widths.
(a) when $t_{1}=t$ and (b) when $t_{1}=1.2t$.}
\end{figure}

Fig.3 shows the relationship between armchair GNR widths ($n$) and
their energy gaps when choosing $t_{1}=t$ and $t_{1}=1.2t$. For
$n=3p$, the energy gap of $t_{1}=1.2t$ is smaller than that of
$t_{1}=t$. This value decreases when the width increases. For
$n=3p+1$, the energy gap of $t_{1}=t$ is zero, which becomes
independent of ribbon width $n$. For $t_{1}=1.2t$,  C-C bonds on the
edge are shortened and the energy gap opens in $(3p+1,0)$ armchair
GNRs. For $n=3p+2$, the energy gaps of $t_{1}=t$ are nearly the same
as that of $t_{1}=1.2t$. By comparing the electronic structure of
$t_{1}=t$ with those of $t_{1}=1.2t$, we observe that the change of
the edge hopping parameter (resulting from the added chemical
groups) can significantly affect the electronic properties of
armchair GNRs. Our calculations reproduce the results based on the local density
approximation of the DFT.\cite{Louie} It is worth mentioning that the energy
gaps of both $t_{1}=t$ and $t_{1}=1.2t$ approach zero, or the effect
resulting from the addends becomes insignificant when $n$ is very
large. This observation suggests that tuning band structure through
edge modifications is effective only for finite-width GNR.

\begin{figure}[htbp]
\includegraphics[width=7cm]{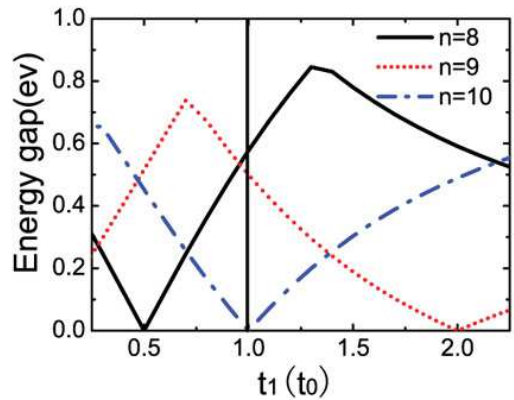}
\caption{(color online) The effect of hopping parameter $t_{1}$ on
energy gaps of armchair GNRs with different widths {[} $n=8, 9$ and
$10$].}
\end{figure}

Next, we focus on evaluating the energy gap as a function of hopping
parameter $t_{1}$ for three kinds of armchair GNRs ($(8,0)$, $(9,0)$,and $(10,0)$).
Results are plotted in Fig. 4. It is interesting to
note that around $t_{1}=t$ (representing the slight deformation
case), the energy gaps of $(8,0)$, $(9,0)$ and $(10,0)$ armchair
GNRs show different trends. In contrast to a $(9,0)$ GNR, the energy
gap of a $(8,0)$ GNR increases as $t_{1}$ increases around
$t_{1}=t$. This trend is also appreciable in Fig.2. A $(10,0)$ GNR
has a zero energy gap at $t_{1}=t$.\cite{PAC90,PRB96,JPSJ96} Once we
applied chemical edge modifications ($t_{1} \neq t$), there is an
opened energy gap that always increases no matter whether $t_1$
decreases ($t_{1}\leq t$) or increases ($t_{1}\geq t$). This
result shows that a transition between metallic and insulating GNRs
is achievable. Moreover, the energy gap can be controlled by
selecting proper addends bounded to the edge carbon atoms of
armchair GNRs.

In Fig.4, one $t_{1}$ value exists that corresponds to the zero
energy gap for both $(8,0)$ and $(9,0)$ armchair GNRs. This
phenomenon shows that, in principle, insulating armchair GNRs can be
modified and become metallic GNRs. However, this kind of
modification is difficult to implement since the hopping parameter
$t_1$ would have to be reduced or enlarged by 100\%.

In conclusion, first-principles calculations show that the chemical
modification of armchair GNRs results in a considerable deformation
of the bond lengths and bond angles near the edge. The introduction
of hopping parameter, $t_{1}$, in the TB scheme accurately
capture the effects caused by chemical edge modifications. Our
theoretical results show that addends can change armchair GNRs' band
structures and even lead to observable metal-to-insulator
transitions. It should be pointed out that the chemical edge
modification is effective only for finite-width GNRs.

\emph{Note}: After submitting this paper, we became aware of the
similar independent work conducted by Son, Cohen, and Louie \emph{et
al.}.\cite{Louie} In their paper,\cite{Louie} the scaling rules for
GNR band gap as a function of their widths and their origins are
studied based on the first-principles approach. A lattice model is
then adopted within TB approximation to explain the changes of energy gap in
armchair GNRs by resolving the Hamiltonian perturbatively. Our
numerical results presented in this paper are obtained by using the exact
diagonalization technique.

This work was partially supported by the National Natural Science
Foundation of China with Grand numbers 10574119, 10674121, 50121202,
and 20533030, by National Key Basic Research Program under Grant No.
2006CB9922000, by the USTC-HP HPC project, and by the SCCAS and
Shanghai Supercomputer Center. Work at NTU is supported in part by
COE-SUG grant (No. M58070001). Jie Chen would like to acknowledge
the funding support from the Discovery program of Natural Sciences
and Engineering Research Council of Canada under Grant No. 245680.
The authors also would like to thank Thomas McIntyre for the
assistance with the finalization of this paper.


\begin{references}
\bibitem{CPL00} C. W. Bauschlicher Jr., Chem. Phys. Lett. \textbf{322}, 237 (2000).

\bibitem{SSP05} M. Burghard, Suf. Sci. Rep. \textbf{58}, 1 (2005).

\bibitem{CR05} X. Lu and Z. F. Chen, Chem. Rev. \textbf{105}, 3643 (2005).

\bibitem{Graphene-Science04} K. S. Novoselov, A. K. Geim, S. V. Morozov, D. Jiang, Y. Zhang,
S. V. Dubonos, I. V. Grigorieva, and A. A. Firsov, Science
\textbf{306}, 666 (2004).

\bibitem{Graphene-Exp} K. S. Novoselov, A. K. Geim, S. V. Morozov, D. Jiang, Nature
\textbf{438}, 197 (2005); Y. B. Zhang , Y. W. Tan , H. L. Stormer ,
Nature \textbf{438}, 201 (2005); C. L. Kane , E. J. Mele , Phys.
Rev. Lett. \textbf{95}, 226801 (2005).

\bibitem{CBSCI06}
Claire Berger, Zhimin Song, Xuebin Li, Xiaosong Wu, Nate Brown,
C\'{e}cile Naud, Didier Mayou, Tianbo Li, Joanna Hass, Alexei N.
Marchenkov, Edward H. Conrad, Phillip N. First and Walt A. de Heer,
Science \textbf{312}, 1191 (2006).

\bibitem{CBJPCB04} C. Berger et al., J. Phys. Chem. B
\textbf{108}, 19912 (2004).

\bibitem{YNASS05}
Y. Niimi, T. Matsui, H. Kambara, K. Tagami, M. Tsukada, and H.
Fukuyama, Appl. Surf. Sci. \textbf{241}, 43 (2005).

\bibitem{YKPRB05}
Y. Kobayashi, K. Fukui, T. Enoki, and K. Kusakabe, Phys. Rev. B
\textbf{71}, 193406 (2005).

\bibitem{YNPRB06}
Y. Niimi, T. Matsui, H. Kambara, K. Tagami, M. Tsukada, and H.
Fukuyama, Phys. Rev. B \textbf{73}, 085421 (2006).

\bibitem{PAC90} H. Hosoya, H. Kumazaki, K. Chida, M. Ohuchi, Y. -D. Gao,
Pure Appl. Chem. \textbf{62}, 445 (1990).

\bibitem{PRB96} K. Nakada,
M. Fujita, G. Dresselhaus, M. S. Dresselhaus, Phys. Rev. B
\textbf{54}, 17954 (1996).

\bibitem{JPSJ96} M. Fujita, K. Wakabayashi, K. Nakada, K. Kusakabe, J.
Phys. Soc. Jpn \textbf{65}, 1920 (1996).

\bibitem{YLPRB02} Y. -L. Lee, Y. -W. Lee, Phys. Rev. B \textbf{66}, 245402
(2002).

\bibitem{9} K. Wakabayashi, M. Fujita, H. Ajiki, M. Sigrist, Phys.
Rev. B \textbf{59}, 8271 (1999).

\bibitem{10} Y. Miyamoto, K. Nakada, M. Fujita, Phys. Rev. B \textbf{59},
9858 (1999).

\bibitem{MolDevices} J. Taylor, M. Brandyge, and K. Stokbro, Phys. Rev. B \textbf{58}, 121101 (2003);
M. Ernzerhof, M. Zhuang, and P. Rocheleau, J. Chem. Phys.
\textbf{123}, 134704 (2005).

\bibitem{vasp1} G. Kresse and J. Hafner, Phys. Rev. B \textbf{49},
14251 (1994).

\bibitem{vasp2}G. Kresse and J. Furthm\"{u}ler, Comput. Mater. Sci.
\textbf{6},15(1996); Phys. Rev. B \textbf{54}, 11169 (1996).

\bibitem{LDA}D. M. Ceperley and B. J. Alder,Phys. Rev. Lett.\textbf{45},
566(1980); J. P. Perdew and A. Zunger, Phys. Rev. B \textbf{23},
5048(1981).

\bibitem{DFT} P. Hohenberg and W. Kohn, Phys. Rev. \textbf{136},
B 864 (1964); W. Kohn and L. J. Sham, \emph{ibid.} \textbf{140},
A1133 (1965).

\bibitem{uspp}D. Vanderbilt, Phys. Rev. B \textbf{41}, 7892 (1990).

\bibitem{Estimation} J. M. Soler, E. Artacho, J. D. Gale, A.Garc\'{\i}a, J. Junquera, P. Ordej\'{o}n, and
D. S\'{a}nchez-Portal J. Phys.: Condens. Matter \textbf{14}, 2745
(2002).

\bibitem{170} D. Porezag, Th. Frauenheim, Th. Kohler, G. Seifert,
and R. Kaschner, Phys. Rev. B \textbf{51}, 12947 (1995).

\bibitem{17} K. Sasaki, S. Murakami, R. Saito, Appl. Phys. Lett.
\textbf{88}, 113110 (2006)

\bibitem{Louie} Y. -W. Son, M. L. Cohen, and S. G. Louie, Phys. Rev.
Lett. \textbf{97}, 216803 (2006).

\end{references}
\end{document}